\documentclass[preprint,3p,times,twocolumn,authoryear]{elsarticle}

\usepackage{amsfonts,amsmath, amsmath,amssymb,slashed}
\usepackage{natbib}
\usepackage{graphicx}
\usepackage{cite,color}
\def\e{{\rm e}}\def\ln{{\rm ln}}

\def\cos{{\rm cos}}

\newcommand{\be}{\begin{equation}}
\newcommand{\bea}{\begin{eqnarray}}

\newcommand{\ee}{\end{equation}}
\newcommand{\eea}{\end{eqnarray}}

\newcommand{\nn}{\nonumber}

\journal{Nuclear Physics B}

\begin{document}

\begin{frontmatter}



\title{Strings from large charged fields}


\author{Carlos A. Cardona}
\ead{cargicar@gmail.com}

\address{IFT-UNESP. R. Dr. Bento Teobaldo Ferraz, 271 - Bloco II, S\~ao Paulo, Brasil}

\begin{abstract}
It has been known long time ago \citet{Polchinski:1996na} that the low energy dynamics of open-strings is described by non-abelian gauge theories in the same way that the low energy effective description of closed strings is governed by Einstein's gravity. In this short note, we review and comment on some examples where conversely, an effective behavior of non-abelian gauge theories in some particular limits are described by some type of extended objects like strings or branes. We  constrain the discussion to a few examples sharing some similarities.

\end{abstract}


\end{frontmatter}


\section{Introduction}
\label{Intro}
This short note is based on a talk given in the X Latin American Symposium of High Energy Physics held in Medellin-Colombia from 24th to 28th of November of 2014, and was intended to address a non-string theory target audience. The main aim is to collect a few  known scenarios where the physics of a particular set of operators in a  field theory, can be effectively be described by string-like or even brane-like configurations at some particular limits.

Stringy behaviors embedded in field theory are not a new topic. A classical example corresponds to vortex in $D=4$ Abelian-Higgs model.  There we can consider static topological solutions confined on two spatial dimension \citet{Abrikosov:1956sx, Nielsen:1973cs}, known as Abrikosov-Nielsen-Olesen vortex, which in three spatial dimensions looks like a rigid string. Those solutions can also be found in supersymmetric extensions where they saturate the BPS condition of the supersymmetric algebra and hence can be protected from quantum corrections, allowing   access to the strong coupling regime of the supersymmetric theory. It is also possible to find an effective action describing the dynamics of those strings.

A related example is given by flux tubes between a quark-antiquark pair in QCD. Lately, there has been a renewed interest on QCD-strings (see for example \citet{Aharony:2013ipa, Shifman:2015kla}) mainly motivated for recent improvements measuring the energies of various states of such strings via lattice simulations (see \citet{Teper:2009uf} for a review and references therein).

Another interesting embedding of strings into gauge theory observables concern to the scattering matrix of massless particles in non-abelian gauge theory. It was noticed by \citet{Witten:2003nn} that the perturbative expansion of ${\cal N}= 4$ super Yang-Mills theory with U(N) gauge group is equivalent to the instanton expansion of a topological string theory whose target space is the Calabi-Yau supermanifold ${\mathbb CP}^{3|4}$. Using this string theory it was possible to reproduce the Maximal Helicity Violating Amplitudes at tree level for gluons in QCD. After this breakthrough and its generalization to include loops \citet{Britto:2004ap}, a huge revolution for the computation of scattering amplitudes in non-abelian gauge theories has started, including the discovery (or confirmation) that a scattering  amplitude of gravitons can be related to a scattering amplitude of gluons. Moreover, the progress in this direction have allowed to apply this techniques to the computation of realistic scattering amplitudes to be search at the LHC (see for example \citet{Deurzen:2014mda}).
 
Along the lines of this note we are going to comment on other stringy  appearances in field theory. A particularity of the chosen examples comes form the similarities into the mechanism how extended objects appear, such as that they usually appear as a collective phenomena for a large charged set of states in the field theory as well as they are associated to non-perturbative effects. 

Despite that we are going to see different examples, we emphasize that we will be bearing in mind the AdS/CFT correspondence \citet{Maldacena:1999pw} relating non-abelian conformal field theories with gravity theories (aka string theories).
Since the original formulation of the correspondence \citet{Maldacena:1999pw}, an impressive amount of work have been done, including many tests of the duality as well as the exact formulation of it for non-abelian conformal field theories in dimensions $D=2,3,4$. It also have been a big deal of progress in the application of the correspondence to more realistic scenarios, such as QCD or Condense Matter Systems. 

\section{Stringy behavior of the Hubbard Model}
Let us start considering a simplified model which will share some characteristics with the gauge theories we are going to consider later on. Let us take a system which transform like a spin-$s$ representation of SU(2) and interact with a external magnetic field $B$ through a Zeeman-like potential $V_z=B\cdot S$, being $S$ the spin-vector $S=(s_x,s_y,s_z)$, \citet{fradkin1991field}. The usual highest weight state representation is defined up to a normalization as the set of states diagonalizing at the same time both $S_z, S^2$,
\be
S_z|s,s\rangle=s|s,s\rangle\,,\quad S^2|s,s\rangle=s(s+1)|s,s\rangle
\ee
The idea is to write the evolution operator of the system in terms of the following set of coherent states,
\be
|n\rangle=\e^{i\theta(n_0\times n)\cdot S}|s,s\rangle\,, 
\ee
where $n_0$ is a unitary vector in the $z$ direction and $\theta$ is the angle between $n_0$ and the arbitrary unitary vector $n$. The above coherent state is an eigenvector of the spin operator $S|n\rangle=s|n\rangle$ and has an expansion in terms of the spins-$s$ representation of SU(2) as,
\be
|n\rangle=\sum_{m=-s}^sD^{(s)}(n)_{m\,s}|s,m\rangle\,, 
\ee
the expansion coefficients satisfies the important property \citet{Perelomov},
\be
D^{(s)}(n_1)D^{(s)}(n_2)=D^{(s)}(n_3)\e^{i\Phi(n_1,n_2,n_3)S_3}\,, 
\ee
where $n_i,\quad i=1,2,3$ are three arbitrary unit vectors and $\Phi(n_1,n_2,n_3)$ is the area enclosed by the spherical triangle with vertices at $n_i$. The normalization of the coherent states is,
\be\langle n_1|n_2\rangle=\e^{i\Phi(n_1,n_2,n_3)s}\left(\frac{1+n_1\cdot n_2}{2}\right)^s\,. \ee
The completes relation in the space of coherent states is given in this case by,
\be 
\mathcal{I}=\left(\frac{2s+1}{4\pi}\right)\int d^3n\delta(n^2-1)|n\rangle \langle n|\,.\ee
The above relations are sufficient to write the evolution operator 
\bea Z&=& {\rm Tr}\e^{i V_z T}\sim\nn\\
&\sim&\lim_{N\to\infty}\left(\prod_{j=1}^{N}\int d\mu(n)\right)\left(\prod_{j=1}^{N}\langle n(t_j)|\e^{-\delta t V_z}|n(t_{j+1})\right)\nn\\
&\sim& \int\mathcal{D}n\,\e^{iS_M[n]}\,,\eea
where $ d\mu(n)=\left(\frac{2s+1}{4\pi}\right)d^3n\,\delta(n^2-1)$ and the action is defined by
\be\label{string} S_M[n]=s\,\mathcal{A}[n]+m\int_0^T dx_0\left(\partial_0n(x_0)\right)^2-s\int_0^T dx_0 B\cdot n(x_0)\,,\ee 
with $m=s\delta t/4\to 0$. The $\mathcal{A}[n]$ corresponds to the area enclosed by the closed trajectory $n(x_0)$ on the unit-sphere surface, and can be rewrote as a Wess-Zumino action on the two-sphere. Hence the action (\ref{string}) can be interpreted as a string on the surface of a sphere whose end point are coupled to a $B-$field. Moreover, it scales with $s$ and hence for large values of $s$, the action is dominated by the classical solutions of the string over the sphere. This stringy description should be replicated in higher dimensions as well as higher symmetrical systems.
\section{Chern-Simons plus matter}
\label{ABJM}
Let us now consider a theory in three dimensions which consist of double Chern-Simons matter theory in three dimension 
with level $(k,-k)$ and gauge group $U(N)\times U(N)$. The theory becomes weakly coupled when the level $k$ is large, hence in the large $N$ limit the coupling analogous to `t Hooft coupling is given by $\lambda\equiv N/k$, which is kept finite. The gauge fields are coupled to four chiral superfields in the bifundamental representation of the gauge group  $U(N)\times U(N)$, and in the fundamental representation of the SU(4) R-symmetry. 
We denote the complex scalars in these 4 chiral multiplets as ($A_1,A_2,\bar{B}_1,\bar{B}_2)$. Here $A_1,A_2$ are in the $(N,\bar N)$ representation of $U(N)\times U(N)$, 
whereas $B_1,B_2$ are in the conjugate, $(\bar N, N)$. Under the SU(4)$_R$ R-symmetry group ($A_1,A_2, \bar{B}_1,\bar{B}_2$) transform in the $\mathbf{4}$
representation.
There is also a $U(1)_R$ under which all of $(A_1,A_2,\bar B_1, \bar B_2)$ have charge $+1$.

We would like to build a particular operator with large charge. In order an operator  to have large charge it need to be composite of a large amount of fundamental fields, but in order to do that, we need to use as a building block an operator in the same $U(N)$, since an index in the left $U(N)$ of the gauge group, cannot be contracted with an index in the right $U(N)$. Without loss of generality, let us take two particular building blocks of that type as
\be Z\equiv(A_2B_2)\,,\,X\equiv(A_2B_1)\,,\ee
which have indexes only on the left $U(N)$ and let us take a particular highly charged operator such as,
\be\label{ojl}
\mathcal{O}_{J,l}=\left[ZZ\cdots ZZ\, X\, ZZ\cdots Z\right]^{m_N}_{p_N}.
\ee
Where the index $J$ denotes the total amount of $Z-$fields in it, i.e, the total charge of the composite operator under a diagonal action of the global symmetry SU(4), and $l$ denotes the position of the ``impurity'' $X$ into the chain of $Z$'s. Denoting by $\Delta$ the dimension of the operator (\ref{ojl}), it satisfies $\Delta-J=1/2$ classically, but this quantity get quantum corrections.  We are particularly interested in computing the first quantum correction to the {\it twist} $\tau\equiv\Delta-J$ of the operator (\ref{ojl}), but even at first order it has a huge mixing whose diagonalization becomes a daunting task. Even so, we would like to do it in a `t Hooft expansion where we need $N$ to be large. It turns out that in that limit, only planar diagrams contribute as the ones displayed in the figure below
\begin{center}
\includegraphics[scale=0.5]{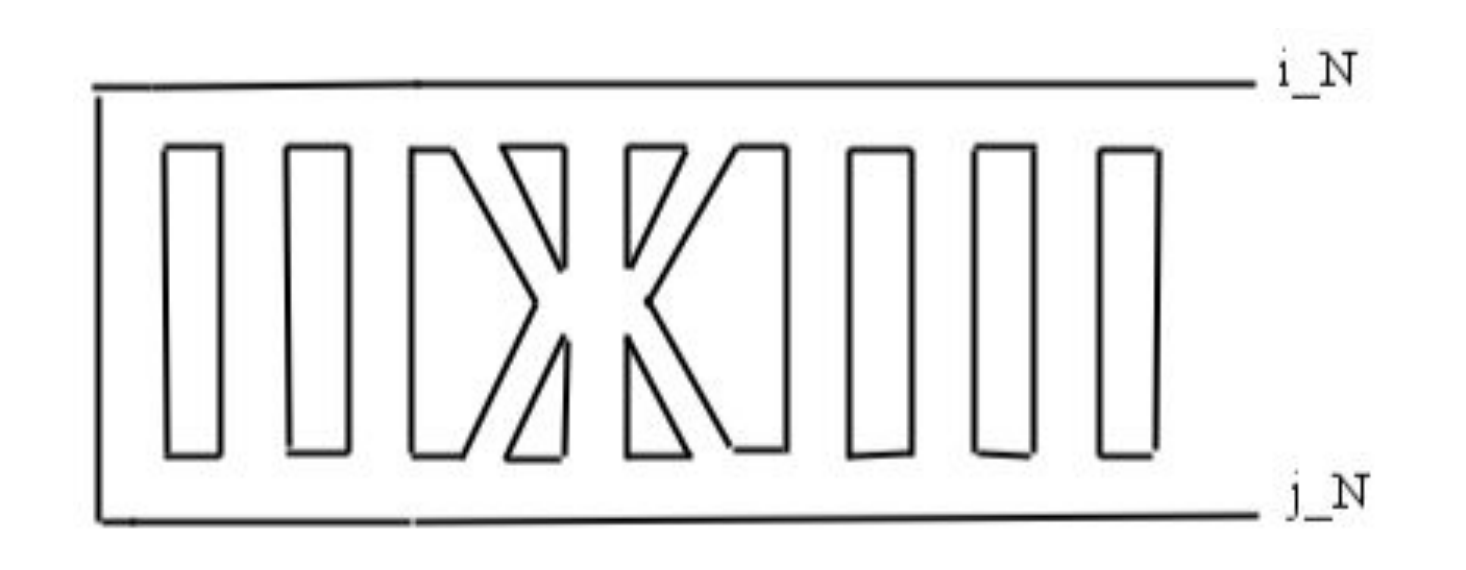}
\end{center}
therefore the operator (\ref{ojl}) with an impurity $X$ at position $l$ only get mixed with operators of the same type with impurities at position $l+1$ and $l-1$. It looks very like a Heisenberg near-neighborhood spin chain, which as is well known can be diagonalized by the following Fourier transformation,
\be\label{ojn}
\mathcal{O}_{J,n}=\times\sum_{l=0}^{J}\left[ZZ\cdots ZZ\, X\, ZZ\cdots Z\right]^{m_N}_{p_N}\cos\left(\frac{\pi n l}{J}\right).
\ee  
The twist of the operator (\ref{ojn}) at leading order is then given by
\be\label{anomalous1}
(\Delta-J_{\chi})=\frac{1}{2}\left[1+\frac{8\pi^2n^2}{J^2}\lambda^2\right]. 
\ee
From the above result, we can see that we could explore the strong coupling ($\lambda>>1$) regime of operators of the type (\ref{ojn}) by taking the limit $J-$large and $\lambda-$large but keeping $\lambda/J$ finite. On the other hand, by doing $J-$large, we could take the continuous limit of (\ref{ojn}), where the chain of $Z$'s become a string and the impurity behaves as a single excitation of the string. By using the $AdS/CFT$ correspondence, it has been proved that it is the actual picture for both, $\mathcal{N}=4$ SYM in four dimensions \citet{Berenstein:2002jq} and for Chern-Simons matter with $\mathcal{N}=6$ in three dimensions \citet{Gaiotto:2008cg} (for open strings in the three-dimensional case see also \citet{Cardona:2014ora}) . In both cases, the large$-\lambda$, large $-J$  limit of the subset of operators just discussed behaves as a string moving really fast on a sphere and on a complex projective space, respectively.
\section{Strings from DIS in QCD}
As a last example, let us consider a more realistic scenario, such as  a process of deeply inelastic scattering DIS of a very energetic hadron off of virtual photon. DIS processes have played a very important role in the understanding of QCD in the early days, in particular it was fundamental in the formulation of the parton model. 

The scattering matrix for an hadron scattered off a virtual photon to an hadron plus a virtual photon, can be written as,
\be
 i\mathcal{M}(\gamma p\to\gamma p)=(ie)^2\epsilon_{\mu}^*(q)\epsilon_{\nu}(q)(-iW^{\mu\nu}(P,q))\,,
\ee 
where $q$ and $\epsilon_{\mu}$ are the momentum and polarization vector of the virtual photon respectively, $P$ is the momentum of the hadron and the matrix
\be 
W^{\mu\nu}=i\int d^4x\e^{iq\cdot x}\langle P|T\{J^{\mu}(x)J^{\nu}(0)\}|P\rangle\,,
\ee
is known as the {\it Forward Compton Amplitude}. 
 $J^{\mu}$ is the quark electromagnetic current 
\be J^{\mu}=\sum_fQ_f\bar{q}_f\gamma^{\mu}q_f\ee
Since we are interested in a very high energetic scattering process, we would like to see at short distances $x\to 0$ and hence we can use the Operator Product Expansion between the electromagnetic currents as follows
\be 
\int d^4x\,\e^{ipx}J_{\mu}(x)J_{\nu}(0)\sim\sum_f\sum_{n=2} c_n^{\mu_1\cdots\mu_{n-2}}{\cal O}^{(n)}_{f\,\mu\nu,\mu_1\cdots\mu_{n-2}}\,,\nn
\ee
where
\be\label{t2}
{\cal O}^{(n)}_{f\,\mu_1\cdots\mu_{n}}=\bar{q}\gamma^{\{\mu_1}D^{\mu_2}\cdots D^{\mu_n\}}q\,,
\ee
Hence the computation is reduced to the expectation value $\langle P|{\cal O}^{(n)}_{f\,\mu_1\cdots\mu_{n}}|P\rangle$, which should satisfies the renormalization group equations (Callan-Symanzik),
\be\label{CZ}
\left(\mu^2\frac{d}{d\mu^2}+\beta(g^2)\frac{\partial}{\partial g^2 }\right){\mathbb O}=-[{\mathbb H}_2(g^2){\mathbb O}]\,,\nn
\ee
Since we would like to focus on high energetic particles, we can neglect the masses of all quarks. In that case, it is convenient to use a ``light-cone formalism" \citet{Kugot, Brink:1982pd, Mandelstam:1982cb} in which one integrates non-propagating degrees of freedom and only keeps the physical ones. It ultimately means that we define every field on a null vector. Calling that vector $n$, it should satisfy that $n^2=1$.

On the light-cone, the evolution kernel appearing in (\ref{CZ}) is given at one-loop by \citet{Balitsky}
\be
 {\mathbb H}_2=\frac{g^2\,C_F}{8\pi^2}[H_{12}+2\gamma_q]\,,
\ee
where $\gamma_q$ is the anomalous dimension of the quark field at one-loop and $H_{12}$ can be represented as an integral operator as 
\bea 
&&[H_{12}\cdot{\mathbb O}](z_1,z_2)=\int_0^1\frac{d\alpha}{\alpha}\bar{\alpha}[2{\mathbb O}(z_1,z_2)\nn\\
&&~~~~~~-{\mathbb O}(\bar{\alpha}z_1+\alpha z_2,z_2)-{\mathbb O}(z_1,\alpha z_1+\bar{\alpha} z_2)]\nn\,,
\eea
where $\bar{\alpha}=1-\alpha$. These operators have the physical interpretation that they displace two particles along the light-cone in the direction of each other.
In order to solve the Callan-Symanzik equation (\ref{CZ}) we should diagonalize ${\mathbb H}_2$. 

Since we are studying a hard processes at large energies, we could use the fact that QCD is approximately conformal at that limit, providing that $\alpha_s\to 0$ \citet{Braun:2003rp}. On the light-cone trajectories, the conformal symmetry reduce to $SL(2,\mathbb{R})$, acting on the coordinate $z$ parametrizing the light-rays as,
\be
 z\rightarrow\frac{az+b}{cz+d},\quad \mathbb{O}(\alpha z)\rightarrow (cz+d)^{-2j}\mathbb{O}\left(\alpha\frac{az+b}{cz+d}\right)\,,
\ee
with $ad-bc=1$ and $j$ being the conformal weight of the field. The infinitesimal generators of the light-cone conformal symmetry $SL(2,\mathbb{R})$ are given by the global Virasoro generators,
\be L^-=-\partial_z,\quad L^+=2jz+z^2\partial_z,\quad L^0=j+z\partial_z\,. \ee
The conformal symmetry tell us that the operators ${\mathbb H}_2$ have to commute with the two particle conformal spin $L_{12}^a\equiv L_1^{a} +L_2^a,\,\,a=\pm,0$ (since the operator (\ref{t2}) is composed of two quarks). It in turn implies that ${\mathbb H}_2$ should be a function of the casimir operator of $SL(2,\mathbb{R})$
\be
 L_{12}^2= L_{12}^a L_{12\,a}=J_{12}(J_{12}-1)\,.
\ee
Acting with the two operators ${\mathbb H}_2$ and $L_{12}^2$ over a representation carrying spin $J_{12}=n+2$  it is found \citet{Korchemsky:2010kj} that,
\be H_{12}=2[\psi(J_{12})-\psi(2)]\,,\ee
where 
\be \psi(x)=\frac{d\ln \Gamma(x)}{dx}\ee
is the Euler psi-function. For large values of $x$ the $\psi(x)$ function can be approximate by,
\be \psi(x)=\ln(x)-\frac{1}{2x}+\mathcal{O}(x^{-2})\,,\ee
therefore, for operators with large spins $J_{12}$ the value under the evolution kernel can be approximate by
\be H_{12}\sim 2\ln J_{12}\,.
\ee
which means that at one loop, the twist $\tau$ of the operator (\ref{t2}) scales as $\ln{J_{12}}$. In the next section, we are going to see that this behavior can be reproduced from a classical string.
\\
{\it Rotating string in $AdS$}
\\
Let us take a spinning closed string in $AdS$ \citet{Gubser:2002tv,Kruczenski:2004wg}. Without loss of generality, let us consider $AdS$ space in three dimensions with the following metric,
\be ds^2=-{\rm cosh}^2 dt^2+d\rho^2+{\rm sinh}^2d\theta^2\,.\ee

We choose an ansatz for a particular rotating string,
\be t=\tau,\quad \theta=\omega\tau+\sigma,\quad \rho=\rho(\sigma)\,.\ee
A classical solution should minimize the Nambu-Goto action on $AdS_3$,
\be
S=-\frac{\sqrt{\lambda}}{2\pi}\int\sqrt{-\dot{X}^2X'^2+(\dot{X}X')^2}\,. 
\ee
By means of the rotating string ansatz  and the action above one can found the equation of motion for $\rho(\sigma)$ given by,
\be \rho'(\sigma)=\frac{1}{2}\frac{{\rm sinh}2\rho}{{\rm sinh}2\rho_0}\frac{\sqrt{{\rm sinh}^22\rho-{\rm sinh}^22\rho_0}}{\sqrt{{\rm cosh}^2\rho-\omega^2{\rm sinh}^2\rho}}\,,\ee
where $\rho_0$ is a integration constant which corresponds to the minimum value of $\rho$ for the given solution. Since we would like to compute the energy and spin of the rotating string, it is not needed to know the exact solution. The quantity of interest is $ E- J$ when $J$ is large. In that limit, the above quantity can be approximated as
\be E-J\sim\sqrt{\lambda}\,\ln(4\pi J)\,.\ee 
In the above expression, $\sqrt{\lambda}$ should be proportional to the string tension, which in the solutions we are considering must be  very big, since the string is strongly stretched when $J$ is large. 
So we see that the classical "twist" $E-J$ of the rotating string in $AdS$ scales with the angular momentum in the same way as the twist of the composite operators appearing at the small-$x$ DIS scattering amplitude.
   
According to the AdS/CFT correspondence, the above rotating string in $AdS$  should corresponds to the strong coupling description of a composite operator with $J-$derivatives in ${\cal N}=4$ SYM.

\section{Conclusions}
In this note we have collected a few known examples of stringy effective descriptions for subsets of large charged operators in non-abelian gauge theories.
We mainly intended to show how extended-like objects appear into field theory, as an effective description of collective phenomena happening at large energies and strong coupling.

In our opinion, it shows how, among other several situations, string theory becomes not only a nice tool to explore interesting limits of field theory but also how it is ubiquitous for a large set of phenomena at high energies.

 {\bf Acknowledgements} 

I would like to thank to the organizer of the X SILAFAE for the hospitality during the event.
The research of CC is supported in part by CNPQ grant 501043/2012-8.




\bibliography{silafaebib}{}
\bibliographystyle{elsarticle-harv} 

\end{document}